# Physical Conditions in Orion's Veil II: A Multi-component Study of the Line of Sight Toward the Trapezium

N. P. Abel[1], G. J. Ferland[1], C. R. O'Dell[2], G. Shaw[1], & T. H. Troland[1]

[1]University of Kentucky, Department of Physics and Astronomy, Lexington, KY 40506; npabel2@uky.edu, gary@pa.uky.edu, troland@pa.uky.edu

[2]Department of Physics and Astronomy, Vanderbilt University, Box 1807-B, Nashville, TN 37235; cr.odell@vanderbilt.edu

## Abstract

Orion's Veil is an absorbing screen that lies along the line of sight to the Orion H II region. It consists of two or more layers of gas that must lie within a few parsecs of the Trapezium cluster. Our previous work considered the Veil as a whole and found that the magnetic field dominates the energetics of the gas in at least one component. Here we use high-resolution STIS UV spectra that resolve the two velocity components in absorption and determine the conditions in each. We derive a volume hydrogen density, 21 cm spin temperature, turbulent velocity, and kinetic temperature, for each. We combine these estimates with magnetic field measurements to find that magnetic energy significantly dominates turbulent and thermal energies in one component, while the other component is close to equipartition between turbulent and magnetic energies. We observe $H_2$ absorption for highly excited $v$, $J$ levels that are photoexcited by the stellar continuum, and detect blueshifted $S^{+2}$ and $P^{+2}$ ions. These ions must arise from ionized gas between the mostly neutral portions of the Veil and the Trapezium and shields the Veil from ionizing radiation. We find that this layer of ionized gas is also responsible for He I 3889Å absorption towards the Veil, which resolves a 40-year-old debate on the origin of He I absorption towards the Trapezium. Finally, we determine that the ionized and mostly atomic layers of the Veil will collide in less than 85,000 years.

## 1 Introduction

Star formation is controlled through a balance among gravitational, magnetic, and thermal energies (e.g. review by Heiles & Crutcher 2005). Magnetic fields are the least understood of these processes because of the few available measurements of field strengths.

Orion's Veil is the absorbing layer of gas in front of the Trapezium region (O'Dell 2001, O'Dell & Yusef-Zadeh 2000) and one of the few regions where *maps of magnetic field strengths exist* (Troland et al. 1989). The Veil is also the location of the anomalously high ratio of total to selective extinction that is common in regions containing rich young star clusters (O'Dell 2002). This paper seeks to understand the role of magnetic fields in Orion's Veil, along with the overall physical conditions in the veil. We discuss the observational data in Section 2, and the model of our first paper in Section 3. In Section 4, we describe the measurement of column densities derived from the analysis of UV spectra. We describe the results of our analysis in Section 5 and summarize our conclusions in Section 6.

## 2 Previous Studies of the Veil

Orion's Veil has a wealth of observational data that makes it suitable for exploring the role of magnetic fields in star-forming regions (see below). Magnetic fields have been measured across the bright central region of M42, but the present study is only concerned with the line of sight towards the Trapezium Cluster, the ionizing stars of M 42.

Our earlier paper, Abel et al. (2004; henceforth A04), analyzed IUE spectra which could not resolve the two main velocity components seen in 21 cm absorption. A04 presented a series of photoionization simulations of the gas and derived global conditions (see Section 3). The current paper briefly reviews the observational data and theoretical framework, then uses higher-resolution UV data to resolve the velocity components seen in 21 cm absorption and to derive physical conditions in each.

### 2.1 H I 21 cm Observations

The background-ionized blister that forms M42 produces strong free-free radio continuum emission, which is absorbed by the Veil. In particular, H I 21 cm absorption has been observed by van der Werf & Goss (1989) and more recently by Troland et al. (in preparation—data given in A04). van der Werf & Goss (1989) identify three $H^0$ velocity components across the face of the Veil, but only two obvious components are associated with the line of sight towards the Trapezium. The VLA observation shows two clear components (see Figure 1). Using the labels of van der Werf & Goss (1989), component A has a $V_{LSR}$ of 1.3 km s$^{-1}$ with a *FWHM* of 3.8 km s$^{-1}$ while for component B these values are 5.3 km s$^{-1}$ and 2.0 km s$^{-1}$ respectively. We report all velocities in the local standard rest (LSR) frame, to convert to heliocentric velocities add 18.1 km s$^{-1}$. Additionally, we use the Doppler *b*-parameter in our analysis of the STIS UV data ($b$ = *FWHM*/1.665).

The 21 cm optical depth profile is shown in Figure 1. This profile, also shown in A04, represents a 25" region centered on $\theta^1$ Ori C and having a velocity resolution of 0.32 km s$^{-1}$. The integrated optical depth for each component is proportional to the ratio $N(H^0)/T_{spin}$, where $T_{spin}$ is the excitation temperature for the 21 cm transition (see, for instance, Spitzer 1978). If the column density is known through other means, such as L$\alpha$ absorption, we can estimate $T_{spin}$ in each component. $T_{spin}$ is often, but not always, approximately equal to the kinetic temperature $T_{kin}$ (Liszt 2001). Component A has $N(H^0)/T_{spin}$ = 1.78×10$^{19}$ cm$^{-2}$ K$^{-1}$ and component B has $N(H^0)/T_{spin}$ = 2.35×10$^{19}$ cm$^{-2}$ K$^{-1}$ (A04).

In the presence of a magnetic field, Zeeman splitting of the $F$=1 state allows the line of sight magnetic field strength for each 21 cm component to be determined. For components A and B, $B_{los}$ = -45 ±5 µG and -54±4 µG, respectively, where the negative sign indicates that the magnetic field vectors are pointed toward the observer. Each of these values is a lower limit upon the total field strength $B_{tot}$. Table 1 summarizes the radio observational data for each component.

## 2.2 Optical

O'Dell & Yusef-Zadeh (2000) combined HST H$\alpha$ and 20 cm radio continuum measurements to measure the extinction across the entire extent of M42 at 1.5" resolution. The Veil is the source of this extinction (O'Dell 2002) and towards the Trapezium has A$_V$ = 1.6 mag.

Optical absorption studies have also yielded information on the chemical composition of the Veil in the directions of the Trapezium and other stars. The most recent and highest resolution study is the one by Price et al. (2001). They observed Ca II, Na I, K I, CH, and CH$^+$ optical absorption lines at many positions in the Veil. A04 argue that Ca$^+$ and Na$^0$ are trace stages of ionization in the Veil owing to the ionizing flux of the Trapezium stars. Hence, the Ca II and Na I lines trace only a small fraction of the mass. We do not consider either line further in our analysis. However, the Price et al. (2001) study did not detect either CH or CH$^+$ in the Veil, with only upper limits of Log[$N$(CH)] ≤ 11.8 and Log[$N$(CH$^+$)] ≤ 12.0. The near absence of UV H$_2$ absorption (see below) means the Veil must be nearly devoid of molecules toward the Trapezium stars. This conclusion is consistent with the absence of detectable CH and CH$^+$ absorption.

## 2.3 UV

The Veil has been observed by three UV missions, Copernicus, the International Ultraviolet Explorer (IUE), and most recently with the Hubble Space Telescope Imaging Spectrograph (STIS). Most of what we know about the chemical composition of the Veil comes from these UV observations.

### 2.3.1 Copernicus

Copernicus was unable to detect H$_2$ in the Veil, setting an upper limit of Log[$N$(H$_2$)] ≤ 17.55 (Savage et al. 1977). In the diffuse ISM, the observed reddening along this line of sight, E(B-V) = 0.32 (Bohlin & Savage, 1981) is usually associated with $H_2$ column densities 2-3 orders of magnitude higher than observed by Copernicus. A04 found that the lack of $H_2$ towards the Trapezium is due to two effects, larger grain sizes found in the Orion environment (Cardelli et al. 1989), and the large UV flux incident upon the Veil owing to its close proximity (~1-3 parsecs) from the Trapezium.

Other regions of the Orion complex are known to contain larger abundance of $H_2$. Recent observations by France and McCandliss (2005) have measured $H_2$ column densities ~$10^{19}$ cm$^{-2}$ for a sightline towards Orion. The sightline they studied, however, was ~12' away from Trapezium in the plane of the sky, and not towards the Trapezium itself.

### 2.3.2 IUE

Most of the column densities for atomic and singly ionized species in the Veil come from the archival study of Shuping & Snow (1997). The spectral resolution of IUE is insufficient to determine column densities for each of the H I 21 cm components, so the column densities they report are the sum of components A and B. Combining their column densities with theoretical calculations, A04 calculated a global model that best fit the observed level of ionization.

Column densities in excited states of C I and O I, and the column density ratios C II/CI, Mg II/Mg I, and S II /S I, were decisive in constraining the photoionization simulations of A04. The excited-state column densities represent the population of the $^3P_{0,1,2}$ states and are referred to as C I, C I*, C I** and O I**, O I*, and O I, respectively. For each, the total column density is just the sum of the column densities of the three levels. The relative populations of these states are sensitive to the hydrogen density, $n_H$, in an $H^0$ region, since collisions with $H^0$ will be the primary excitation mechanism for these levels (Keenan 1989; Roueff & Le Bourlot 1990). The column densities of Mg I & II, S I & II, and C I & II tell us about the level of ionization for a given element. By reproducing the observed levels of ionization, the A04 model was able to place the Veil ~1-3 parsecs from the Trapezium.

### 2.3.3 STIS

The most recent and highest resolution UV spectra of the line of sight to $\theta^1$ Ori B are presented in Cartledge et al. (2001, 2005; GO 8273) and Sofia et al. (2004; GO 9465). Cartledge et al. used the E140H echelle grating centered at 1271 Å, while Sofia et al. used the E230H grating centered at 2263, 2313, and 2363 Å. These studies dealt primarily with total abundances along many lines of sight in the galaxy, so they only reported the total column density of O I, Kr I, C II Mg II, P II, Ge II, Ni II, and Cu II. Cartledge et al. (2001) also derive a total H I column density towards $\theta^1$ Ori B from Lα absorption. Towards this line of sight, $N(H^0)$ = 4.8 (± 1.1) ×$10^{21}$ cm$^{-2}$. For comparison, Shuping & Snow (1997) analyzed IUE observations to find $N(H^0)$ = 4.0 (± 1.0) ×$10^{21}$ cm$^{-2}$ toward $\theta^1$ Ori C.

The STIS velocity resolution of ~2.5 km s$^{-1}$ allows comparison between the 21 cm H I and UV absorption lines. Independent of the 21 cm analysis, Cartledge et al. (2001) found O I absorption lines with $V_{LSR}$ of 0.9 and 5.9 km s$^{-1}$, which correlate well with the 21 cm data. The *b* values reported in their absorption model, 1.9 (±0.4) and 1.3 (±0.7) km s$^{-1}$, are similar to the 21 cm *b* values of 2.3 (±0.04) and 1.2 (±0.02) km s$^{-1}$. These values for *b* were derived from the Kr I line, and they should be entirely due to turbulence because of the large mass of Kr.

The correspondence in *V* and *b* between the UV and radio data makes it extremely likely that both profiles are sampling the same gas components. It should be possible to determine column densities for many species in each velocity component independently. For comparison, Figure 1 shows many of the absorption lines used in this analysis on the same velocity scale as the 21 cm H I profile.

# 3 The A04 Model

The previous work of A04 determined a "best-fit" model to the observational data given in section 2 (the STIS spectra, however, was not available at the time of A04). Our calculations were performed with the spectral synthesis code Cloudy, last described by Ferland et al. (1998). We considered a constant density, plane parallel slab illuminated on one side by the Trapezium. We use gas-phase abundances observed in the Orion Nebula (Rubin et al. 1991; Osterbrock et al 1992, Baldwin et al. 1991). A few of the abundances by number are He/H = 0.095, C/H = 3.0$x$10$^{-4}$, N/H = 7.0 $x$10$^{-5}$, O/H=4.0 $x$10$^{-4}$, Ne/H=6.0$x$10$^{-5}$ and Ar/H =3.0 $x$10$^{-6}$. However, the lightest thirty elements are included in our models. The stellar ionizing continuum is the modified Kurucz LTE atmosphere described by Rubin et al. (1991), which was modified to reproduce high-ionization lines seen in the H II region. We set the total number of ionizing photons emitting by $\theta^1$ Ori C equal to 10$^{49.34}$ s$^{-1}$, a typical value for an O6 star (Osterbrock & Ferland 2005).

Our calculations were non-equilibrium but time steady. The velocities and linewidths of the H I 21 cm components suggest that component A and B are quiescent regions not recently affected by shocks. The time scale for H$_2$ formation of ~5×10$^8$/$n_H$ yrs (Bertoldi & Draine 1996) combined with the $n_H$ derived in A04 means that time-dependent effects should be damped out over a timescale of 2-5×10$^5$ yrs. When compared to the estimated age of the star cluster, 10$^6$ yrs (O'Dell 2001), we conclude that the Veil is likely in steady-state equilibrium, and justifies the use of a steady state calculation.

## 3.1 Grain Physics

The A04 model used the improved grain physics described by van Hoof et al. (2004; see also Abel et al. 2005) which explicitly solves for the grain temperature and charge as a function of grain size. We divide the grain size distribution into ten size bins, and include grains that are composed of graphite and silicates. Our size distribution is designed to reproduce the flat UV extinction curve observed by Cardelli et al. (1989) towards Orion. We used an $A_V/N(H_{tot})$ ratio of $3.5 \times 10^{-22}$ mag cm², which is ~65% of that seen in the general ISM, but consistent with the values of Av and $N(H^0)$ derived in the literature (see Sections 2.2 and 2.3). We scaled our grain abundance to match this ratio.

## 3.2 H$_2$ Physics

The microphysics of H$_2$ in Cloudy is described in Shaw et al. (2005). Energies and radiative transition probabilities for the 301 ro-vibrational levels within the ground electronic state $1s\ ^1\Sigma_g$ (denoted as X) are taken from Stancil (2002, private communication) and Wolnievicz et al. (1998). We have included the ro-vibrational levels within the lowest 6 electronic excited states that are coupled to the ground electronic state by permitted electronic transitions. Energies and radiative transition probabilities for excited electronic states are from Abgrall, Roueff, & Drira (2000). These electronic excited states are important because they determine the Solomon process, which destroys H$_2$ through the absorption of Lyman-Werner band photons from the ground electronic state followed by decays into the X continuum. These photo-excitations are also an indirect source of population of excited ro-vibration levels within X which decay to produce infrared emission lines with the selection rule $\delta J = 0, \pm 2$. We have taken photoionization cross sections for transitions into the continuum of the Lyman and Werner bands from Allison & Dalgarno (1969). Effects of cosmic rays and the x-ray continuum are included as well. H$_2$ can be formed either on dust grains in a cold dusty environment or from H$^-$ in a hot dust-free environment. The state specific (*v* and *J* resolved) formation rates of H$_2$ on grain surfaces and via H$^-$ route have been taken from Cazaux & Tielens (2002); Takahashi, Junko, & Uehara (2001); and Launay, Dourneuf, & Zeippen (1991). Line overlap and self-shielding are also considered. In addition, our calculations are designed to reproduce the observationally determined H$_2$ formation rate on grains in the diffuse ISM, determined by Jura (1974) to be ~$3x10^{-17}$ cm$^{-3}$ s$^{-1}$.

## 4 Data & Analysis

Five STIS spectra are in the HST archives. This paper focuses on the single short-wavelength E140H spectrum with its many atomic, ionic and molecular lines. The longer wavelength E230H spectra include the C II] 2325Å line analyzed by Sofia et al. (2004) and a wealth of Fe II lines. We do not consider the E230H spectra in this study.

## 4.1 Data reduction

We used two software packages to analyze the STIS data. We began with the pipeline processed (using CALSTIS 2.17b) x1d.fits files and used IRAF[1] (Tody, 1993) tasks to improve the S/N by coadding overlapping spectra from different echelle orders and grating tilts. The resulting average S/N ratio is 30-35, as determined by IRAF. The continuum flattened, normalized spectra were then analyzed with the spectral fitting package VPFIT[2] to derive column densities. VPFIT employs a $\chi^2$ analysis to fit Voigt profiles to one or more velocity components in absorption lines. For each such component, VPFIT accepts an initial guess for the column density (*N*), heliocentric center velocity (*V*) and line width (*b*). VPFIT then determines the best-fit profile to an absorption line and reports estimates for *N*, *V* and *b*. VPFIT also estimates errors in these quantities based upon the IRAF-derived S/N ratio of the data.

Several atomic and instrumental parameters are needed for this analysis. Rest wavelengths $\lambda_{ij}$ and oscillator strengths $f_{ij}$ for each transition come from Morton (2003), except for $H_2$ transitions where we use data from Abgrall et al. (2000) and Wolnievicz et al. (1998). The instrumental response profile, which is convolved with the Voigt profile, is also required. We assume the instrumental profile for the E140H echelle grating is a Gaussian with a FWHM of 1.37 km s$^{-1}$ (STIS Handbook).

We used VPFIT on the STIS spectra to derive column densities $N_A$, $N_B$ for each of the two Veil velocity components, A and B. In doing so, we fixed the center velocities $V_A$, $V_B$ of the two components to the values derived very accurately from the 21 cm H I absorption profile (Figure 1, Table 1). The fitting process involves an estimate of the Doppler width parameter *b* for each component. These parameters cannot be well determined because the instrumental response profile provides only about seven independent spectral channels across an absorption line. However, column densities for the two velocity components are insensitive to *b* for lines of relatively low optical depth. Therefore, we were able to derive accurate column densities for the two velocity components for many species in the Veil. These include C I, C I*, C I**, O I, Cu II, Ni II, Ge II, P II, and Mg II (Table 2). In Table 1 we list values for the *b* parameters derived by VPFIT for the optically thin lines of O I] λ1355.57Å, Cu II λ1358.77Å, and Kr I λ1235.84 Å. Additionally, we also show the fit to the O I] line in Figure 2. As expected, the *b* values have rather large errors. The column densities listed in Table 2 are systematically ~0.1 dex smaller than those reported by Cartledge et al. (2005) and is primarily due to differences in data calibration (Stefan Cartledge, private communication).

---

[1] IRAF is distributed by the National Optical Astronomy Observatories, which are operated by the Association of Universities for Research in Astronomy, Inc., under cooperative agreement with the National Science Foundation

[2] VPFIT is available at www.ast.cam.ac.uk/~rfc/vpfit.html

For optically thick (saturated) lines in the STIS spectrum, it is impossible to derive $N_A$, $N_B$ independently. For these lines, we derived the *total* column density for all components of a given species along the line of sight. Optically thick lines for which total $N$ were derived include S II, S III, N I, Si II, Si II*, O I*, O I**, P III, and C II. Optical spectra reveal other absorption lines along this line of sight (Price et al., 2001). However, they have very different velocities and are clearly not associated with components A or B. These extra components, however, could affect the analysis of the broad, saturated lines. The error introduced in not doing a multi-component fit to the saturated lines is ~1 dex, and therefore we believe the column densities of the saturated lines in Table 2 are only correct to within an order of magnitude.

## 4.2 Column Densities

We can estimate $N(H^0)$ for each velocity component from STIS data. The Lα line (Cartledge et al. 2001) yields total $N(H^0)$. From species that are expected to be the dominant ionization stage in $H^0$ regions, we can apportion $N(H^0)$ between the two velocity components. The $H^0$ region should be fully filled with $O^0$ and $Kr^0$ since these elements have ionization potentials > 13.6 eV. Also, Kr is not depleted in the ISM (Cartledge et al. 2001), so a constant Kr/H ratio is expected for both components. We find that 2/3 of the Kr column density resides in component B, and the same is true for O. Therefore, the same ratio must apply to $H^0$, and we conclude that $N(H^0) \sim 1.6 \times 10^{21}$ cm$^{-2}$ for component A and $\sim 3.2 \times 10^{21}$ cm$^{-2}$ in component B (Table 3).

We have measured excited-state column densities for both components A and B. These include level populations for C I ($^3P_0$), C I* ($^3P_1$), and C I** ($^3P_2$). Through the theoretical calculations given in A04, we then determine the volume density in each velocity component. In another useful density diagnostic, O I* and O I**, both lines become somewhat saturated. However, the derived velocity of O I* and O I** absorption (~0 km s$^{-1}$) makes it likely that both excited states come from component B. Also, as we will show below, the density implied C I* absorption is consistent with O I* and O I** coming from a denser component B.

In addition to the atomic and first ion species, the E140H spectrum shows H$_2$ in absorption. This is the first reported H$_2$ column density along the line of sight towards the Trapezium. Table 4 summarizes our results. Previous missions failed to detect H$_2$ owing to a lack of spectral resolution and sensitivity (IUE, Copernicus) or the inability to observe the trapezium (FUSE). The observed transitions are the L0-4 $P(3)$ 1342.26 Å, L0-3 $P(3)$ 1283.11 Å, L0-3 $R(3)$ 1278.72 Å, and L0-4 $R(1)$ 1333.8 Å, where the usual spectroscopic notation has been used. The $v$=3-0 $P(3)$ and $v$=3-0 $R(3)$ H$_2$ absorption lines arise from the same lower level; therefore, the four absorption lines can be used to derive column densities for three H$_2$ levels. A comparison between the H I and H$_2$ $v$=3-0 $P$(3) line profiles is shown in Figure 1.

We measured the equivalent width ($W_\lambda$) of each transition and each component. We then determined the column density by assuming the absorption is on the linear part of the curve of growth. This assumption should be valid, since the optical depth for all the observed lines was 0.2 or smaller. We find that $N$(H$_2$) for component A is at least five times smaller than $N$(H$_2$) for component B (Table 4).

## 4.3 Identification of a Blue-shifted Velocity Component in [N II] 6583 Å

The [N II] 6583 Å emission line data of Doi, O'Dell, & Hartigan (2004) was used to search for velocity components not associated with the primary layer of ionized material that provides most of the light from M42. In Figure 3 we show an average spectrum of a region 45.5"x110.7" (long axis oriented north-south) centered 50.7" with a position angle of 132° from θ$^1$ Ori C. This data set has a FWHM of about 8 km s$^{-1}$. The asymmetric line profile has been deconvolved with IRAF subtasks into three components, all having a FWHM of 16 km s$^{-1}$. The strongest component at 0 km s$^{-1}$ has 81% of the total energy. To the red (16 km s$^{-1}$) is a component with 6% of the total energy, while to the blue (-19 km s$^{-1}$) is a component with 13% of the total energy. The strongest component arises from the primary layer of emission and the redward component must be the velocity shifted light arising from scattering in the background PDR beyond M42's ionization front (O'Dell, Walter & Dufour 1992, Henney 1998, O'Dell 2001).

Using the method of calibration of the HST WFPC2 emission line filters (O'Dell & Doi 1999) and the images of O'Dell & Wong (1996), one finds that the total [N II] surface brightness is 2.6×10$^{-13}$ erg cm$^{-2}$ s$^{-1}$ arcsec$^{-2}$, with an error of ± 20%. The average extinction at H$\beta$ in the sample region is about $c$(H$\beta$) = 0.6 (O'Dell & Yusef-Zadeh 2000), so the extinction correction at 6583 Å is a factor of 2.69. The extinction corrected surface brightness is therefore 7.0×10$^{-13}$ erg cm$^{-2}$ s$^{-1}$ arcsec$^{-2}$. This corresponds to a surface brightness of 9.2×10$^{-14}$ erg cm$^{-2}$ s$^{-1}$ arcsec$^{-2}$ for the blueward component. Although Doi, O'Dell, & Hartigan (2004) show that there are blue-shifted velocity components across the face of M42 due to high velocity shocks, they are too weak to account for the component we see in this large-area sample. As described in Section 5.3, the blueward component is an important constraint to the ionized gas.

# 5 Results

## 5.1 The Density & Temperature of the Neutral Layers

The populations of levels within ground terms of atoms depend on density and temperature (Jenkins et al. 1998). Their observed column densities are then related to physical conditions. For component A the observed ratio of excited to total carbon is $N(C\ I^*)/N(C\ I_{tot}) = 0.34$ with a range between 0.29 and 0.76 allowed, and $N(C\ I^{**})/N(C\ I_{tot}) = 0.13$ with an allowed range of 0.03 to 0.39. For component B $N(C\ I^*)/N(C\ I_{tot}) = 0.43$ with a range of 0.32 to 0.55 and $N(C\ I^{**})/N(C\ I_{tot}) = 0.36$ ranging from 0.26 to 0.48 respectively. Such high values are only possible in regions where $n_H$, $T$, or some combination of $n_H$ and $T$, is high (Keenan 1989). Similar ratios can be found in our data set for O I*, O I**, and Si II* column densities, but the lines are saturated so we cannot resolve both components.

Several estimates are possible for temperatures in the two velocity components. We can estimate $T_{spin}$ in each component by combining $N(H^0)$ (Table 3 and Section 4.2) with the ratio of $N(H^0)/T_{spin}$ derived from the 21 cm data. We find $80 < T_{spin} < 110$ K for component A and $100 < T_{spin} < 165$ K for component B, with most probable values of 90 K and 135 K, respectively. Information about $T_{kin}$ comes from two sources. First, we can place upper limits to $T_{kin}$ by assuming the 21 cm H I line widths are entirely due to thermal motions. These limits are $T_{kin} < 87$ K for component A and $T_{kin} < 320$ K for component B. Second, we can use the density-temperature relationship found by A04. The densities derived in this section (see below) imply $T_{kin} = 50$ K for component A and $T_{kin} = 80$ K for component B.

These estimates suggest $T_{kin} < T_{spin}$ for both components. In many astrophysical environments, $T_{kin} = T_{spin}$. This equality holds as long as the level populations that determine $T_{spin}$ are in thermal equilibrium (Liszt 2001). We plan to discuss reasons why $T_{kin} < T_{spin}$ in a future paper.

These observed excited state column density ratios, the predicted values presented in Figure 4 of A04, and the predicted $T_{kin}$ allow $n_H$ to be determined. The range of the level population ratios quoted above either measures a density or places a lower limit to the density. For component A we find $n_H \approx 10^{2.5}$ cm$^{-3}$ with the range $10^{2.1} < n_H < 10^{3.7}$ cm$^{-3}$ allowed, while for component B $n_H \sim 10^{3.4}$ cm$^{-3}$ with $n_H > 10^{2.3}$ cm$^{-3}$. We can also use the observed $N(C\ I^*)/N(C\ I_{tot})$ and $N(C\ I^{**})/N(C\ I_{tot})$ ratios to simultaneously solve for $n_H$ and $T_{kin}$ in each component. Such an approach only changes the derived density by less than a factor of two, which is much less than the uncertainty in $n_H$ that comes from the 1σ range of each column density ratio. The Copernicus upper limit to $N(H_2)$, combined with Figure 6 of A04, requires that $n_H < 10^4$ cm$^{-3}$ in both components. In Section 5.2, we show that the H$_2$ column density is, to within 1σ, between $10^{14}$ - $10^{15}$ cm$^{-2}$. This H$_2$ column density further constrains $n_H$ to be between $10^2$ and $10^{3.5}$ cm$^{-3}$ for both components.

Our estimates for $N(H^0)$ and $n_H$ for each component can be used to derive the corresponding physical thickness $l$ of each region ~1.7 parsecs for component A and ~0.4 parsecs for component B. The average density of both components is $\overline{n_H} = N(H^0)/l = 10^{2.9}$ cm$^{-3}$, very close to A04's value of $10^{3.1\pm0.2}$ cm$^{-3}$ derived from IUE observations that did not resolve the lines.

The O I lines are saturated and cannot be deconvolved into two components. The measured total column density ratios are $N(O\ I^*)/N(O\ I_{tot}) \sim 10^{-3.5}$ and $N(O\ I^{**})/N(O\ I_{tot}) \sim 10^{-3.8}$. This ratio is an increasing function of density (A04) so almost all O I* and O I** absorption comes from the dense component B. The predicted O I* and O I** is almost an order of magnitude smaller for the lower density of component A.

## 5.2 H$_2$ Abundances in Both Components

The detection of H$_2$ further constrains the physical conditions in the Veil. The observed H$_2$ absorption lines arise from highly excited $v=3, 4$ and $J=1, 3$ levels, a signature of rapid photodissociation through excited electronic states (Black & van Dishoeck 1987). To illustrate this effect, we recomputed the best-fitting model, using the values of $n_H$ and $N(H^0)$ given in Table 4 and including the H$_2$ model described in Shaw et al. (2005). Figure 4 shows the predicted column densities as functions of $v$ and $J$.

For this plot, the higher $J$ values at a given $v$ correspond to higher excitation energies. The calculation shows that the level populations peak for the $J=1, 3$ ortho-H$_2$ levels, which are also the ones observed in the STIS spectrum. We only detect levels from $v = 3$ and 4 because of the available wavelength coverage. Lower-$v$ levels have higher population but they produce lines shortward of our STIS spectrum.

The $v$ and $J$ H$_2$ states we see in absorption do not trace the bulk of H$_2$. The absorption lines detected are strong lines that lie in unblended regions of the spectrum. Other $v$ and $J$ states with higher H$_2$ column densities are either weaker or blended. Strong lines are predicted for $\lambda < 1215$ Å but the spectrum has less signal to noise in this region. Other lines are blended with C II 1335 Å, Lα, or other strong lines.

The levels we detect have very high excitation potentials, between 1.7-2.2×10$^4$ K above the ground state. A04 show that the Veil is predominantly atomic because of the rapid H$_2$ photodissociation rate. This Solomon process (Abgrall et al. 1992) mainly leads to population of highly excited levels within the ground electronic state, producing the populations we observed. The level populations for H$_2$ are therefore far from thermodynamic equilibrium because of UV pumping (Black & van Dishoeck 1987).

Even though the observed levels do not trace the majority of $H_2$, we can estimate the total column density by combining Figure 4 with the measured column densities. We made this estimate by multiplying $N[H_2(v,J)]$ derived from each line, by the ratio $N[H_2(total)]/N[H_2(v,J)]$ predicted by our models and given in Table 4. The predicted total column density, the last column of Table 4, then allows both the errors in the observations and the models to be judged. The observational errors are substantial and we include the same level, measured from two lines, to judge this error. We find the total $N(H_2)$, summed over both components, to be $\approx 10^{14.6}$ cm$^{-2}$ with an error of 0.3 dex.

The $H_2$ column density in the Veil is close to the point where self-shielding by the Lyman/Werner bands decreases the $H_2$ destruction rate (Draine & Bertoldi 1996). Because each component moves with a different radial velocity, neither component can shield the other. Additionally, for atomic regions with a high $H_2$ destruction rate, $N(H_2)$ will scale linearly with both $n_H$ and $N(H)$ (Draine and Bertoldi 1996). Both $n_H$ and $N(H)$ are larger for component B, which explains why the $H_2$ column density is higher in that component. Given the low $H_2$ abundance in the Veil, it is unlikely that other molecules will be observable for this sightline.

## 5.3 The H$^+$ Region near the Veil

STIS data show that an H$^+$ region exists somewhere in *front* of the Trapezium, as S III and P III absorption lines must come from an H$^+$ region. This H$^+$ region is distinct from the main H$^+$ region that lies behind the Trapezium stars. The observed S$^{2+}$ column density is within 0.1 dex of the A04 model (Table 2 of A04), while the P$^{2+}$ differs from A04 by 0.35 to 0.91 dex. Note that A04 used the observed S abundance for the Orion H$^+$ region while the P abundance has not, to our knowledge, been measured. If P$^{2+}$ is the dominant stage of ionization, then this implies that the P/H ratio in Orion is ~5×10$^{-7}$, which is within 25% of the observed solar abundance (Savage & Sembach 1996). The value of 1.6×10$^{-7}$ used in A04 may, therefore, be ~0.5 dex too small. If S$^{2+}$ is the dominant stage of ionization in the H$^+$ region, then this implies that $N(H^+)$ for the ionized gas is ~10$^{20}$ cm$^{-2}$.

In addition to the STIS data, optical studies of the Veil also reveal the presence of an H$^+$ layer. The major difficulty in deconvolving the contributions from various regions is that the entire range of velocities seen along the line of sight, ~20 km s$^{-1}$, is not much larger than the thermal line width for 10$^4$ K gas. The [N II] 6583Å emission (A04) and the He I 3889Å absorption (O'Dell et al. 1993) must come from ionized gas. The observed [N II] surface brightness of 9.2×10$^{-14}$ erg cm$^{-2}$ s$^{-1}$ arcsec$^{-2}$ agrees within 25% of the value predicted by A04 (7.3×10$^{-14}$ erg cm$^{-2}$ s$^{-1}$ arcsec$^{-2}$), and the predicted optical depth of He I 3889Å (0.85), which A04 did not report, is within 15% of the observed value, $\tau(3889) \sim 1$ (O'Dell et al. 1993).

The S III, P III, He I, and [N II] lines formed in the H$^+$ layer all have roughly the same velocity, indicating these species are located in the same physical region. Figure 1 shows S III absorption on the same velocity scale as the 21 cm H I absorption. The lines are quite optically thick, accounting for the noise near the line center. The line center velocities are -13.6 ± 0.9 km s$^{-1}$ for S III and -13.3 ± 3.0 km s$^{-1}$ for P III. The large error is due to line saturation for S III, and because the P III absorption line is blended with a C II absorption line. O'Dell et al. (1993), derived a velocity of -17.1 ± 1.0 km s$^{-1}$ for the He I 3889Å absorption line, which is consistent with the velocity of the [N II] 6583Å emission line (-19.0 ± 3.0 km s$^{-1}$, see Section 4.3). O'Dell et al. (1993) also reported that the unpublished thesis of Jones (1992) showed an [O II] velocity component at -15 km s$^{-1}$ near the Trapezium.

The velocity of the H$^+$ layer is surprising if it is the ionized surface of the Veil facing the Trapezium, as suggested by the agreement with predictions of A04. We know that an H$^+$ region must lie between the Trapezium and the Veil in order to provide shielding for the atomic layers. If the H$^+$ region is a photoevaporative flow from the main Veil then the ionized gas is accelerated towards the source of ionization in the rest frame of the Veil. However, the H$^+$ velocities are observed to be approaching the observer at ~-15 km s$^{-1}$ in this rest frame. If the ionized gas lies between the Trapezium and the Veil then it must be a matter-bounded layer which happens to be moving towards the Veil. If the ionized layer is a photoevaporative flow then it must lie on the near side of the Veil. In this case, we do not detect the H$^+$ layer that shields the Veil from the ionizing radiation of the Trapezium and would be forced to conclude that the Veil is considerably further from the Trapezium than was deduced in A04.

The H$^+$ layer is in one of two possible configurations relative to the Veil; between the Trapezium and Veil or between the Veil and Earth (see A04 for the geometry of this line of sight). If the H$^+$ layer is on the far side of the Veil, the layer is ionized by the Trapezium, primarily $\theta^1$ Ori C. If the H$^+$ layer is on the near side of the Veil, then the source of ionization must be ι Ori, an O9 III star 300-600 parsecs from earth (Gualandris, Portegies Zwart, & Eggleton 2004) and 31′ from the Trapezium in the plane of the sky. This was the placement favored by O'Dell et al. (1993). However, the star ι Ori is not luminous enough to account for the ionization. The [N II] surface brightness given in Section 4.3 and Figure 8 of A04 is roughly independent of density when the density is below the line's critical density. In this case, the surface brightness is set by the flux of ionizing photons striking the layer. If $\theta^1$ Ori C ionizes the H$^+$ layer, then the layer's distance from $\theta^1$ Ori C is ~1.3 parsecs (A04). Since an O9 star has ~10 times fewer ionizing photons than an O6 star (Vacca et al. 1996), ι Ori would have to be ~0.4 parsecs away from the H$^+$ layer in order to explain the [N II] surface brightness. The angular separation of ι Ori from the line of sight to the Trapezium, the minimum distance ι Ori can be from the H$^+$ layer, is 4.4 parsecs and its true distance must be larger. This conclusively places the H⁺ layer between the Veil and Trapezium.

Our identification of the H⁺ layer resolves a 40-year controversy over the physical model for the Orion Nebula (for a complete description of the controversy, see the review by O'Dell (2001)). The presence and strength of the He I absorption 3889Å line has been known since the work of Adams (1944). Wurm (1961) used the strength and blue-shift of this line (with respect to the He I emission lines) to argue that the Orion Nebula is a thin layer of ionized material lying beyond the Trapezium stars. This is the correct model usually credited to Zuckerman (1972) and Balick, Gammon, & Hjellming (1973). Münch & Wilson (1962) applied a very different interpretation to the same data, arguing that the nebular gas was optically thick to its own dust component (proven incorrect in O'Dell (2002)) and distributed symmetrically about the Trapezium stars. The newly discovered H⁺ region reconciles, for the first time, the correct physical model for the nebula, the line strength, and its velocity.

We conclude that the H⁺ layer is a distinct matter-bounded layer that lies between the Trapezium and the Veil, and happens to be moving towards the Veil. The fact that an ionized layer is not detected at the Veil velocity suggests that the H⁺ layer we detect extinguishes much of the ionizing radiation produced by the Trapezium. Therefore the layer has a significant optical depth in the Lyman continuum. We have computed the radiative acceleration of the gas as part of the evaluation of the radiation – gas interactions (Henney et al. 2005). We find that the H⁺ layer has a radiative acceleration $a_{\rm rad}$ equal to $5\times10^{-7}$ cm s$^{-2}$ and is due to Lyman continuum absorption by hydrogen and, to a lesser extent, by dust. The separation between the stars and the Veil is ~2 pc. A layer of gas that is free to move over this distance would be accelerated to a velocity of 25 km s$^{-1}$ in ~$1.5\times10^5$ years.

Dynamical instabilities may provide an explanation for why the H⁺ layer absorbs nearly all of the ionizing radiation. Mathews (1982) points out that a radiatively accelerated photoionized cloud becomes Rayleigh-Taylor unstable at the H⁺ ionization front and he proposed that this circumstance provides a natural process to truncate a cloud at this point – neutral gas is not efficiently accelerated and ablates off the H⁺ zone.

The H⁺ layer is an extended layer that covers much of the same region as the Veil. He I absorption is seen towards all Trapezium stars (O'Dell et al. 1993), and also towards θ² Ori A, which is ~2′ away from the Trapezium in the plane of the sky. At a distance of 500 parsecs, this shows that the gas has a lateral extent of at least 0.25 pc in the plane of the sky. If the gas has the same density as the Veil (~$10^3$ cm$^{-3}$) then it has a physical thickness of 0.1 parsecs. Like the layers of the Veil, it appears as a sheet of gas.

We know, however, that the H⁺ layer and Veil do not interact and therefore are separated by some distance. The A04 model, which had the H⁺ layer as a simple extension of the mostly atomic gas, placed the Veil ~2 parsecs away from the Trapezium. This calculation under-predicts the observed [N II] surface brightness by ~20%. To within the observational error, Figure 8 of A04 places the ionized layer 1.1-1.6 parsecs away from the Trapezium, with a mean value of 1.3 parsecs.

We can estimate the collision time between the $H^+$ and $H^0$ layers. A04 found that the 1σ estimate of the $H^0$ layers from the Trapezium is 1-3 parsecs (A04). We also have a 1σ estimate for the distance of the $H^+$ layer. The maximum separation distance between the two layers is therefore 1.9 parsecs. Given the $H^+$ layer's current radiative acceleration and relative velocity (15 km s$^{-1}$), a collision between the $H^+$ layer and the Veil will happen in < 85,000 years. If we take the best-fit values for the distance of the $H^+$ and $H^0$ layers from the Trapezium, then the collision time is ~40,000 years.

## 5.4 Energetics

Physical parameters in the two Veil layers reveal the relative importance of thermal, turbulent, magnetic and even gravitational energies. These energies can be estimated from the observed linewidths ($\Delta v_{tot}$), magnetic field strengths ($B_{los}$) and column densities $N(H^0)$, together with model values for the gas density ($n_H$) and $T_{kin}$ (Table 1 and 3). Moreover, energies in the two Veil layers may be compared with typical values for the galactic cold neutral medium (CNM), an interstellar regime that is also observable in 21 cm H I absorption. We use the best-fit model values for $n_H$ and $T_{kin}$ to compute the most probable energies for each H I velocity component. We also compute each parameter using upper and lower limits upon $n_H$ and $T_{kin}$ as established by the model or by other constraints. For parameters involving the magnetic field, we apply a statistical correction to convert the observed $B_{los}$ into the most probable estimate for the total field strength $B_{tot}$. These factors are $B_{tot} = 2B_{los}$ and $B^2_{tot} = 3B^2_{los}$, appropriate for parameters dependent upon $B$ and $B^2$, respectively (Crutcher 1999). Definitions and symbols for physical parameters below (e.g. the turbulent Mach number $M_{turb}$) are taken from Heiles & Troland (2005). In converting $n_H$ in the mass volume density, we have assumed the usual fractional helium abundance of 0.1.

In Table 5 we list the derived parameters for the two Veil components along with median values for these parameters appropriate to the CNM of the galaxy. These latter values come from the 21 cm H I absorption survey of Heiles & Troland (2005 and references therein). For each parameter, we list the most probable value along with a range of values. The ranges are based upon the allowed ranges in $T_{kin}$ and $n_H$ in the two Veil components (Section 5.1). The turbulent Mach number $M_{turb}$ *squared* is the ratio of turbulent to thermal energy densities. Values of $M_{turb} > 1$ imply supersonic turbulence. Veil component B is very supersonically turbulent, like the CNM. Veil component A appears more quiescent, it may be only mildly supersonic. The parameter $\beta_{therm}$ is the ratio of thermal to magnetic energy densities. This ratio is always < 1 in the interstellar medium and often << 1 in self-gravitating regions (Crutcher 1999, who uses the symbol $\beta_p$). For the Veil components, $\beta_{therm} \ll 1$. The parameter $\beta_{turb}$ is the ratio of turbulent to magnetic energy densities. This same ratio is also expressed in terms of the Alfvenic turbulent Mach number, where $M^2_{Alf,turb} = 0.67\ \beta_{turb}$. In self-gravitating clouds, existing magnetic field observations suggest $\beta_{turb} \approx 1$, a state of magnetic equipartition (See Crutcher 1999, who provides values for the Alfvenic Mach number $m_A$, where $m_A^2 = \beta_{turb}$). Also, in the non self-gravitating CNM, $\beta_{turb} \approx 1$, again a state of approximate magnetic equipartition. However, Veil component A has $\beta_{turb} \ll 1$. Therefore, turbulent energy densities in this component are much less than magnetic energy densities. The energetics of Veil component A appear strongly dominated by the magnetic field, the first such region in the interstellar medium yet to be identified. This conclusion is consistent with the results of A04 who modeled values for $n_H$ and $T_{kin}$ averaged over the two Veil components. Note that A04 characterized $\beta_{turb}$ in terms of the ratio of the actual gas density $n_H$ to the equipartition density $n_{eq}$. The latter density is the approximate density required for magnetic equipartition, and $n_H/n_{eq} = 1.9\beta_{turb}$. Veil component B has $\beta_{turb} \approx 1$, consistent with magnetic equipartition as in the CNM and in self-gravitating clouds. In summary, Veil component B is similar to CNM and to most self-gravitating gas in the galaxy in that thermal energy is insignificant compared to turbulent and magnetic energies, and the latter two are comparable (magnetic equipartition). However, Veil component A is quite different. In this component, magnetic energy strongly dominates thermal and turbulent energies, both of which are comparable although relatively small.

    Another magnetic parameter of interest is the mass-to-flux ratio, M/Θ. This parameter is a measure of the ratio of gravitational to magnetic energies in a *self-gravitating* cloud. A critical value exists for this ratio such that a self-gravitating cloud will be supported indefinitely by magnetic pressure if M/Θ for the cloud is less than the critical value and if slippage between charged and neutral particles (ambipolar diffusion) does not occur. (See Crutcher 1999 and references therein.) Of course, the Veil layers of Orion are not likely to be self-gravitating. However, the ratio M/Θ is conserved if ambipolar diffusion is not an important process. Therefore, M/Θ in the Veil today may be representative of M/Θ in Orion gas that once was (or will become) self-gravitating. Crutcher (1999) provides a relationship for λ, the ratio of M/Θ to the critical value. In particular, $\lambda = 0.5 \times 10^{-20}\ n(H^0)/B$, where $n(H^0)$ is in cm$^{-3}$ and $B$ is in μG. Applying this relation and using $B = B_{los}$ for each component (Table 1), we find that λ = 0.18 and 0.30 for components A and B, respectively. These values of λ are *upper limits* because $B \geq B_{los}$. We conclude that λ < 1 for both Veil components, so they both are said to be magnetically subcritical. The CNM also appears to also be magnetically

subcritical (Table 5).  However, self-gravitating gas generally appears to be magnetically critical ($\lambda \approx 1$, Crutcher 1999).

## 5.5 Veil properties

Some properties of the Veil resemble the CNM in the galaxy.  These include kinetic temperature, existence of supersonic turbulence and a magnetically subcritical state.  In other ways, of course, the Veil is quite different, reflecting, no doubt, its association with OMC-1.  $H^0$ column densities in each of the Veil components are an order of magnitude greater than typical CNM column densities of $2-3 \times 10^{20}$ cm$^{-2}$ (Heiles & Troland 2005).  Also, $n_H$ is 1-2 orders of magnitude higher in the Veil than in typical CNM, and the magnetic field strengths are an order of magnitude higher.  As a result, total pressures in the Veil components are comparable to each other and about two orders of magnitude higher than in the CNM or in diffuse interstellar material in general.  The Veil layers are not likely to be gravitationally stable on their own.   They may be confined by the gravitational field of the background molecular cloud OMC-1, coupled to the cloud via the magnetic field.  However, higher spatial resolution extinction observations (O'Dell & Yusef-Zadeh 2000) show that there are small clumps in the Dark Bay region of the Veil.

While distinctly different in many physical properties from the CNM, the two Veil components are also different from each other.  Component A appears less supersonically turbulent, somewhat cooler, significantly lower in density, and thicker than component B.  In addition, component A is dominated by magnetic field energy.  Curiously, magnetic field strengths in both components seem essentially the same despite the factor of eight lower density in component A implied by the best fit model.  That is, there is no apparent connection of field strength to gas density in the two Veil components, a property that also applies to diffuse gas in the galaxy (e.g. Heiles and Crutcher 2005).   Both the Veil components and the CNM are magnetically subcritical.  It is possible, therefore, that interstellar gas in a magnetically subcritical state does not experience an increase in field strength with density.  This phenomenon, which bears further theoretical study, may result from motion of gas primarily along field lines or else from decoupling of the neutral gas from the field, that is, ambipolar diffusion.

Which of the two Veil components lies closer to the Trapezium?  Van der Werf & Goss (1989) argued that component B arises in dissociated $H_2$ gas, closer to the stars while component A represents halo gas from the molecular cloud, undisturbed by the Orion $H^+$ region.  However, the $H_2$ abundance in both components can be explained without considering the Veil-Trapezium distance.  Additionally, the $H^+$ region near the veil (Section 5.3) is decoupled from the Veil gas. We therefore cannot definitively say which component is closer to the Trapezium.

# 6 Conclusions

We have combined archival STIS high resolution data with 21 cm data to analyze the region known as Orion's Veil. Our analysis tells us that:

1. The UV and 21 cm observations have two primary velocity components, a narrower component A and a broader component B.

2. $n_H$ is ~$10^{2.5}$ cm$^{-3}$ for component A and ~$10^{3.4}$ cm$^{-3}$ for component B. About 2/3 of the total H$^0$ column density is associated with the broader of the two observed 21 cm absorption components. $N(H^0)$ is derived from column densities of species that must coexist with H$^0$, so we can determine $T_{spin}$ for each component. It is likely that $T_{kin} \leq T_{spin}$ in both components. The physical thickness of component A is ~1.3 parsecs and ~0.5 parsecs for component B.

3. We observe, for the first time towards the Trapezium, H$_2$ in absorption. We find that component A has fewer molecules than component B, due to a lower $n_H$ and H$^0$ column density. The H$_2$ absorption lines emerge from highly excited states, consistent with the Veil being relatively close to the Trapezium.

4. We detect an ionized layer of gas along the line of sight to the Trapezium. The S III, P III, and He I absorption lines all have similar velocities, indicating they are associated with the same region that produces an [N II] emission line component of the nebular spectrum.

5. The surface brightness of the newly identified blueshifted [N II] places the ionized gas between the Trapezium and the Veil. The velocity of the ionized gas is blueshifted with respect to the Veil, indicating that it is a distinct layer moving towards the Veil. We find the column density of the H$^+$ layer is ~$10^{20}$ cm$^{-2}$, and it is located 1.1-1.6 parsecs away from the Trapezium. Our characterization of this H$^+$ region resolves a 40-year-old debate on the origin and strength of the He I 3889Å absorption line.

6. The ionized layer is probably a radiatively accelerated sheet pushed outward by radiation pressure. At its current distance, the ionized gas will collide with the Veil in < 85,000 years.

7. The energy in component A is dominated by the magnetic field, a situation unique in the ISM. In component B, magnetic and turbulent energies are in approximate equipartition as they appear to be in most regions of the ISM. Both components A and B are magnetically subcritical.

8. We cannot determine, given the current data, if component A or B is closer to the Trapezium. Higher resolution H$_2$ absorption data with better S/N is needed to determine strength of the UV radiation field and place each Veil component along the line of sight.

Acknowledgements: We would like to thank many people for useful discussion related to this work. In particular, we would like to acknowledge Blair Savage for his many invaluable comments on drafts of the paper. We would also like tot thank R. Srianand, Stefan Cartledge, and Dan Welty for useful discussions related to our data analysis. This work was funded through HST program 10636 and NSF AST 03-07720. N.PA. would like to acknowledge the support of the Kentucky Space Grant Consortium and the Center for Computational Sciences at the University of Kentucky. T.H. T. acknowledges the support of NSF grant 03-07642.

# 8 Tables

Table 1 – Observational Data

| Parameter | Component A | Component B |
|---|---|---|
| $N(H^0) / T_{spin}$ (cm$^{-2}$ K$^{-1}$) | 1.78×10$^{19}$ | 2.35×10$^{19}$ |
| $b$ - H I (km s$^{-1}$) | 1.20 ± 0.02 | 2.30 ± 0.04 |
| $V_{LSR}$ (km s$^{-1}$) | 5.30 ± 0.01 | 1.30 ± 0.03 |
| $b$ - O I] λ1355.57Å (km s$^{-1}$) | 1.56 ± 0.43 | 2.20 ± 0.32 |
| $b$ - Cu II λ1358.77Å (km s$^{-1}$) | 1.61 ± 0.35 | 2.05 ± 0.28 |
| $b$ - Kr I λ1235.84 Å (km s$^{-1}$) | 1.72 ± 1.01 | 2.14 ± 0.76 |
| $B_{los}$ (µG) | -45 | -54 |

Table 2 – STIS Derived Column Densities in the Veil

| Species | Rest Wavelength of Transitions (Å) | Log[N(X)] (cm$^{-2}$) Component A | Log[N(X)] (cm$^{-2}$) Component B | Log[N(X)] (cm$^{-2}$) Total |
|---|---|---|---|---|
| O I | 1355.60 | 17.51 ± 0.06 | 17.83 ± 0.04 | 18.00 ± 0.03 |
| C I ($^3P_0$) | 1260.48, 1277.25, 1280.14, 1328.83 | 12.86 ± 0.13 | 12.86 ± 0.13 | 13.16 ± 0.09 |
| C I* | 1260.93, 1261.00, 1261.12, 1276.75, 1277.28, 1277.51, 1279.89, 1280.40, 1280.60, 1329.08, 1329.10, 1329.12 | 12.67 ± 0.24 | 13.19 ± 0.10 | 13.30 ± 0.10 |
| C I** | 1261.43, 1261.55, 1277.19, 1277.55, 1277.72, 1280.33, 1280.85, 1329.58, 1329.60 | 12.24 ± 0.47 | 13.12 ± 0.10 | 13.17 ± 0.10 |
| C I$_{tot}$ | ………………………… | 13.14 ± 0.22 | 13.56 ± 0.10 | 13.69 ± 0.06 |
| Ni II | 1317.22 | 13.12 ± 0.03 | 13.40 ± 0.02 | 13.58 ± 0.02 |
| Cu II | 1358.77 | 12.28 ± 0.05 | 12.52 ± 0.04 | 12.72 ± 0.03 |
| P II | 1301.87 | 13.71 ± 0.33 | 14.32 ± 0.24 | 14.41 ± 0.20 |
| Kr I | 1235.84 | 12.08 ± 0.11 | 12.31 ± 0.08 | 12.51 ± 0.06 |
| Ge II | 1237.06 | 11.79 ± 0.05 | 11.95 ± 0.04 | 12.17 ± 0.03 |
| Mg II | 1239.93, 1240.39 | 15.30 ± 0.04 | 15.58 ± 0.03 | 15.76 ± 0.02 |
| C II[1] | 2325.40 | N/A | N/A | 17.82 (+0.12, -0.19) |
| O I* | 1304.86 | N/A | N/A | 14.31[2] |
| O I** | 1306.03 | N/A | N/A | 14.06 |
| N I | 1199.55, 1200.22, 1200.71 | N/A | N/A | 17.74 |
| Si II | 1304.37, 1260.42, 1193.29, 1190.42 | N/A | N/A | 16.52 |
| Si II* | 1309.28, 1265.00, 1264.74, 1197.39, 1194.50 | N/A | N/A | 15.37 |
| P III | 1334.81 | N/A | N/A | 13.59 |
| S II | 1259.52, 1253.80, 1250.58 | N/A | N/A | 15.97 |
| S III | 1190.20 | N/A | N/A | 14.98 |

[1] Column density taken from Sofia et al. (2004)

[2] Lines without a reported error are saturated, and the error is of the order 1 dex

Table 3 – Physical Conditions

| Parameter | Component A | Component B |
|---|---|---|
| $N(H^0)$ cm$^{-2}$ | 1.6(1.4-2.0)×10$^{21}$ | 3.2(2.4-3.9)×10$^{21}$ |
| $n_H$ cm$^{-3}$ | 10$^{2.5}$ (10$^{2.1}$-10$^{3.5}$) | 10$^{3.4}$ (10$^{2.3}$-10$^{3.5}$) |
| $l$ (parsecs) | 1.3 | 0.5 |
| $T_{spin}$ (K) | 90 (80-110) | 135 (100-165) |
| $T_{kin}$ (K) | 50 (< 87) | 80 (< 80) |

Table 4 – H$_2$ Absorption Data

| H$_2$ Absorption Line | Component A | | | | Component B | | | |
|---|---|---|---|---|---|---|---|---|
| | W$_\lambda$ (mÅ) | $N$(H$_2$)$_{obs}$ (cm$^{-2}$) | [H$_2$(tot)/H$_2$($v,J$)]$_{A04}$ | [$N$(H$_2$)]$_{tot}$ (cm$^{-2}$) | W$_\lambda$ (mÅ) | $N$(H$_2$)$_{obs}$ (cm$^{-2}$) | [H$_2$(tot)/H$_2$($v,J$)]$_{A04}$ | $N$(H$_2$)$_{tot}$ (cm$^{-2}$) |
| L0-3 $R$(3) 1278.72 | 1.58 ± 0.60 | $10^{12.27\ +0.14}_{\ \ \ \ \ -0.21}$ | 63.4 | $10^{14.1}$ | 2.28 $^{+1.25}_{-0.81}$ | $10^{12.49 \pm 0.19}$ | 73.8 | $10^{14.4}$ |
| L0-3 $P$(3) 1283.11 | 0.56 ± 0.47 | $10^{11.82\ +0.26}_{\ \ \ \ \ -0.81}$ | 63.4 | $10^{13.6}$ | 3.44 $^{+0.51}_{-0.44}$ | $10^{12.77 \pm 0.06}$ | 73.8 | $10^{14.6}$ |
| L0-4 $R$(1) 1333.80 | 0.51 ± 0.70 | $10^{11.78}$ (<12.15) | 135.8 | $10^{13.9}$ | 2.50 $^{+2.27}_{-1.19}$ | $10^{12.28 \pm 0.28}$ | 158.2 | $10^{14.5}$ |
| L0-4 $P$(3) 1342.26 | 0.39 ± 0.36 | $10^{11.67\ +0.28}_{\ \ \ \ \ -1.03}$ | 97.8 | $10^{13.7}$ | 2.82 $^{+0.49}_{-0.44}$ | $10^{12.52 \pm 0.07}$ | 113.1 | $10^{14.6}$ |
| | | | $\overline{N(\mathrm{H_2})_{tot}}=$ | $10^{13.8}$ | | | $\overline{N(\mathrm{H_2})_{tot}}=$ | $10^{14.5}$ |

Table 5 - Magnetic Parameters for the Veil

| Parameter | Component A | Component B | CNM |
|---|---|---|---|
| $M_{turb}$ | 1.8 (0 – 2.2) | 3.5 (3.4 – 5.0) | 3.7 |
| $V_A$ (km s$^{-1}$)[1] | 9.4 (3.0 – 14.8) | 4.0 (3.6 – 14.2) | 1.5 |
| $\beta_{therm}$ | 0.01 (0.003 – 0.17) | 0.088 (0.004 – 0.12) | 0.29 |
| $\beta_{turb}$ | 0.014 (0.007 – 0.014) | 0.50 (0.045 – 0.6) | 1.9 |
| $\lambda$ | 0.09 (0.08 – 0.11) | 0.15 (0.11 – 0.18) | 0.22 |

---

[1] $V_A$ is the Alfven velocity

A

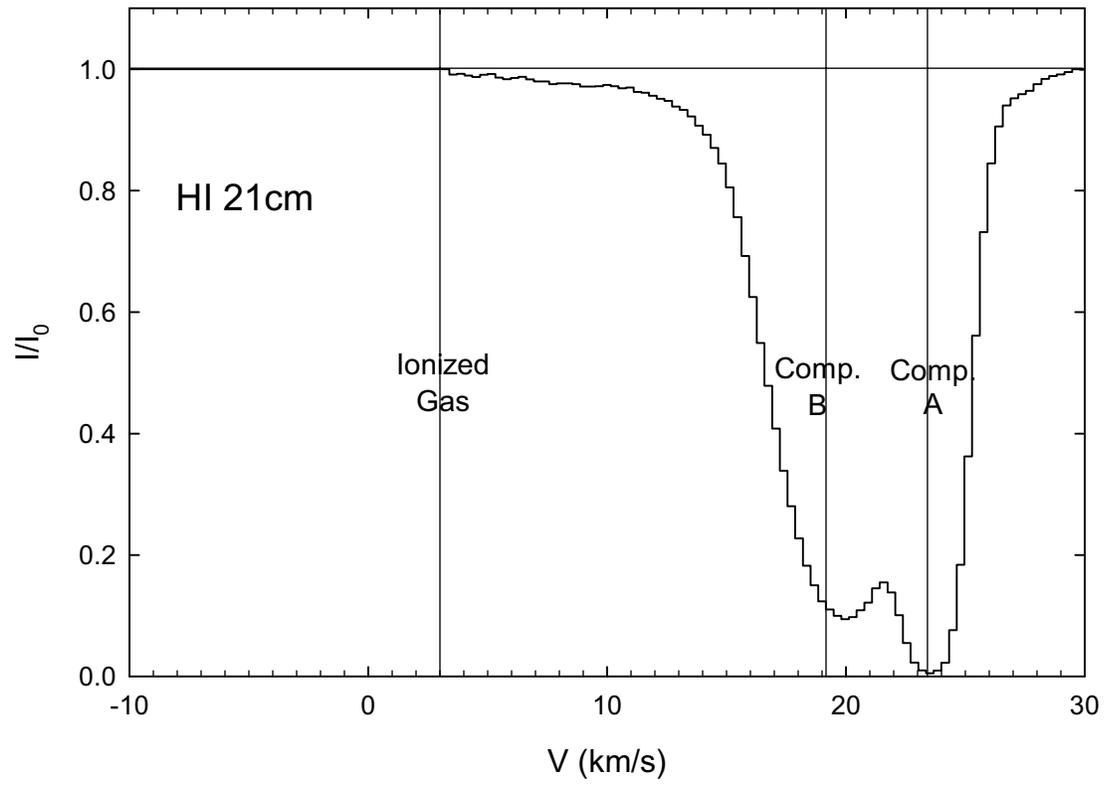

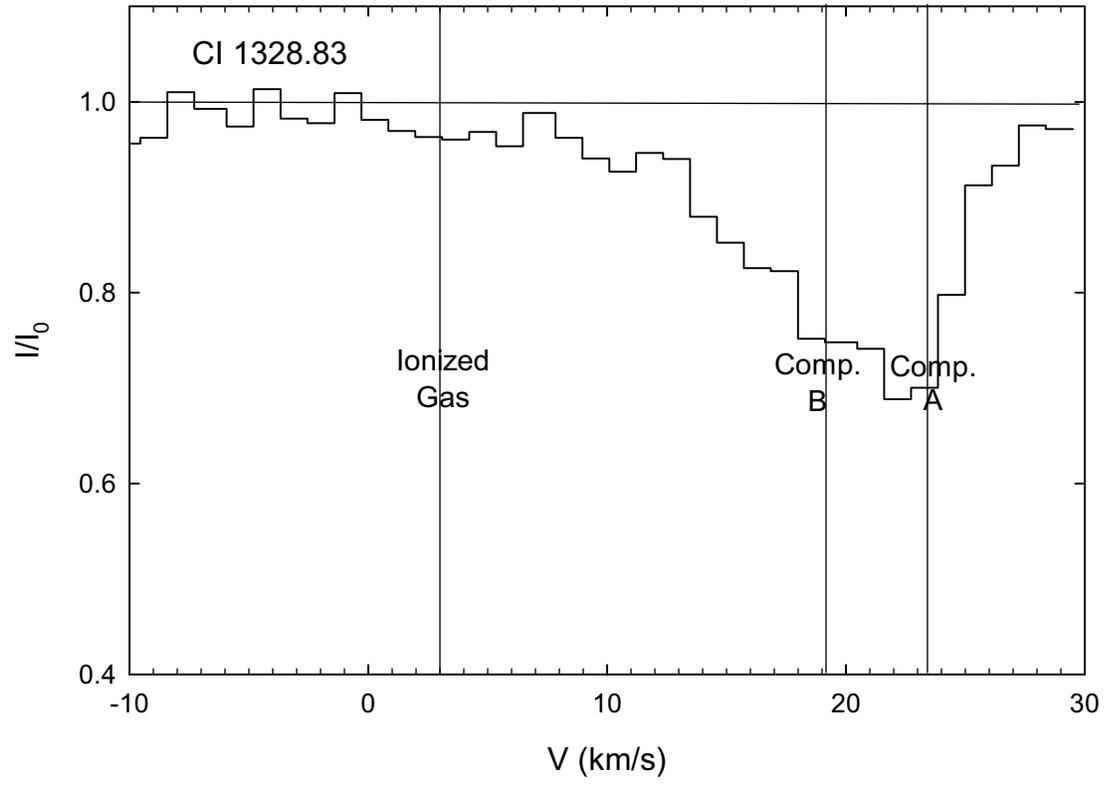

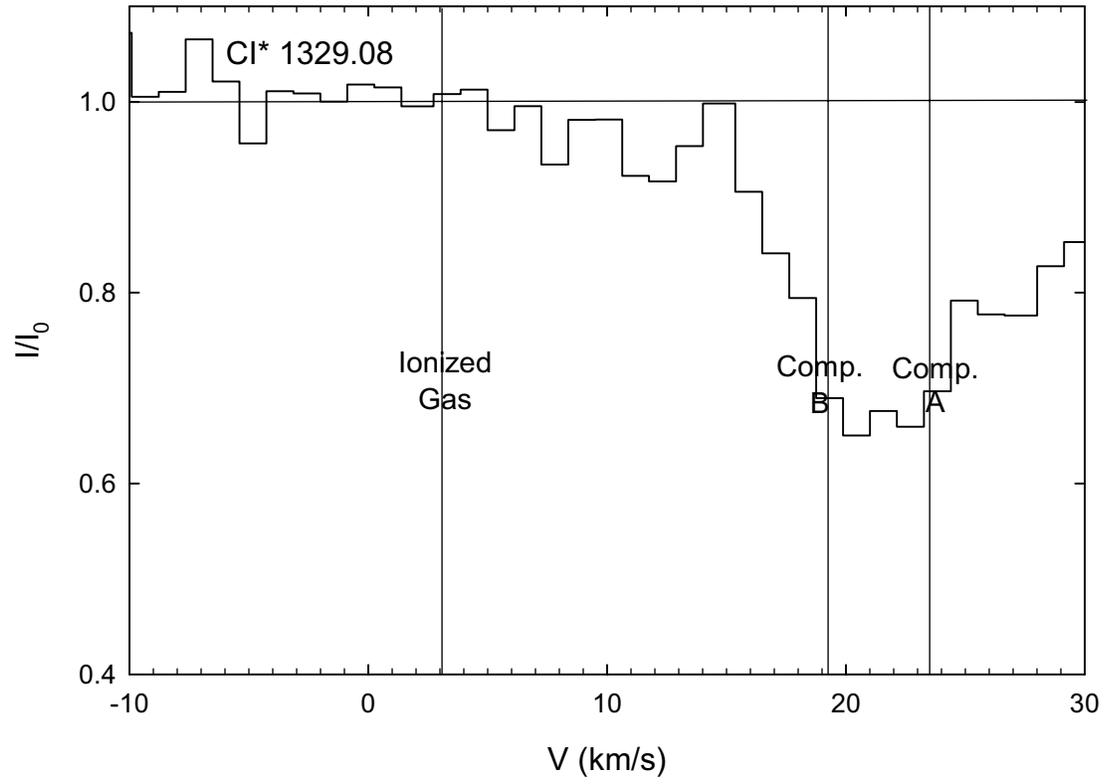

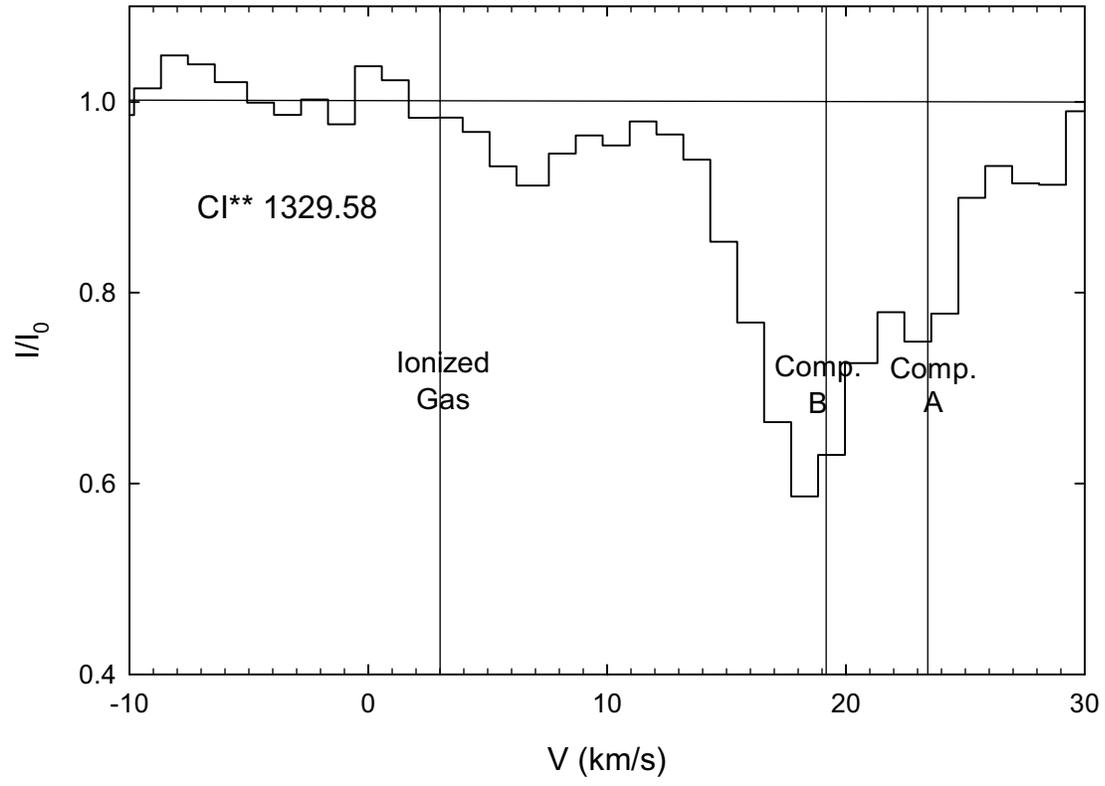

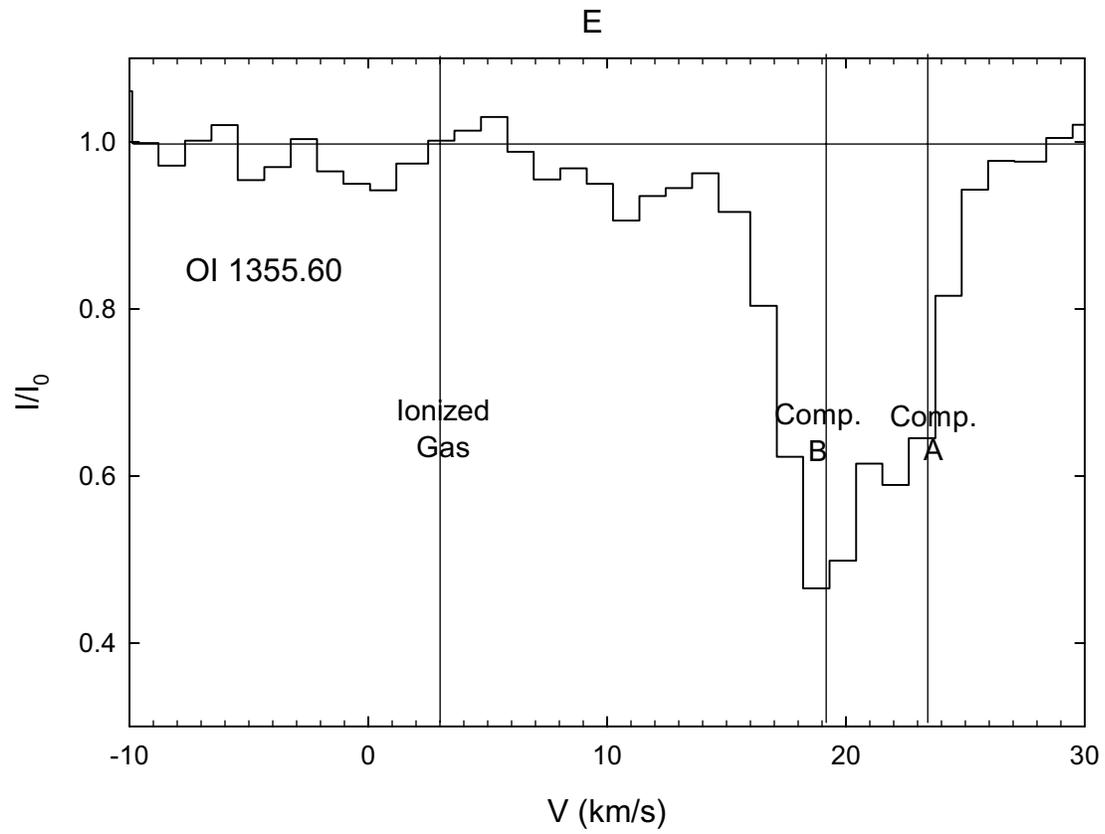

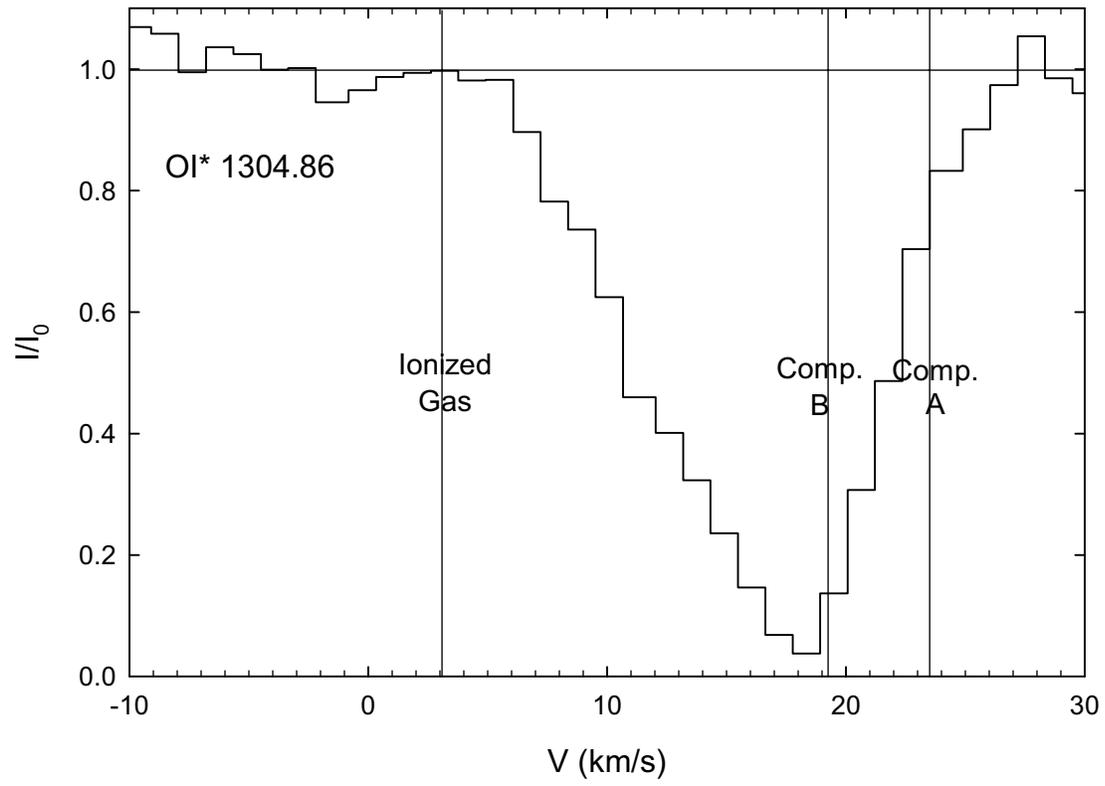

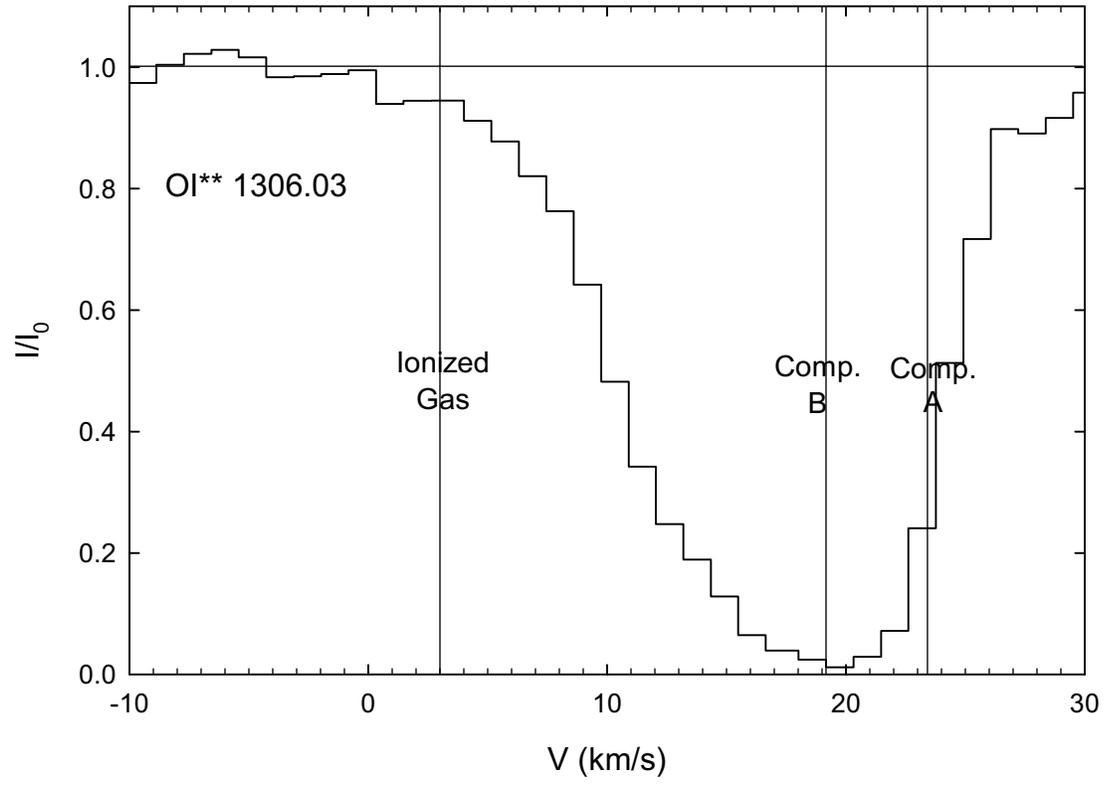

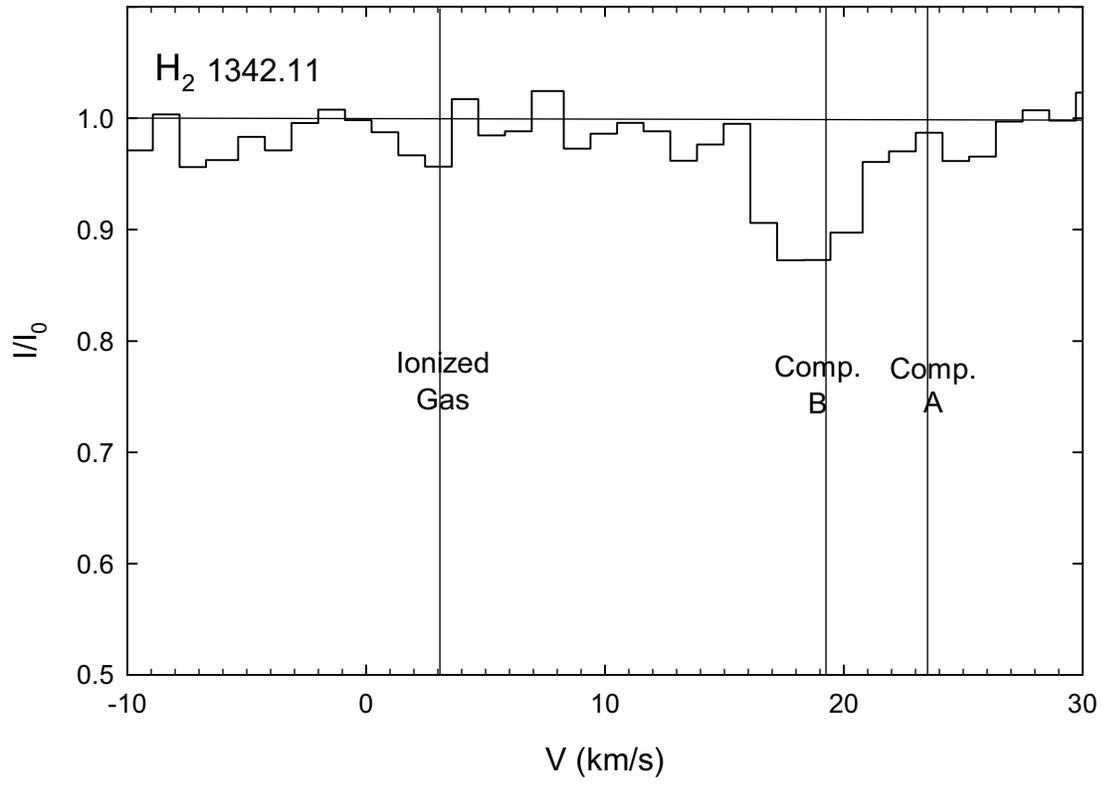

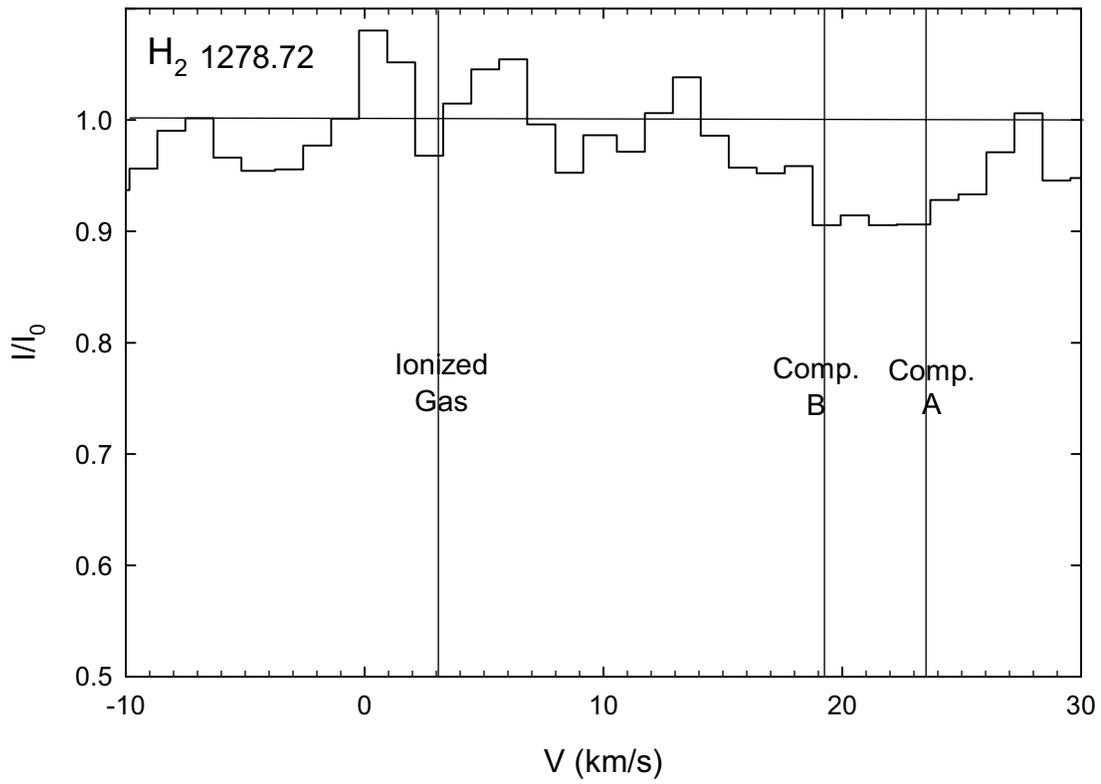

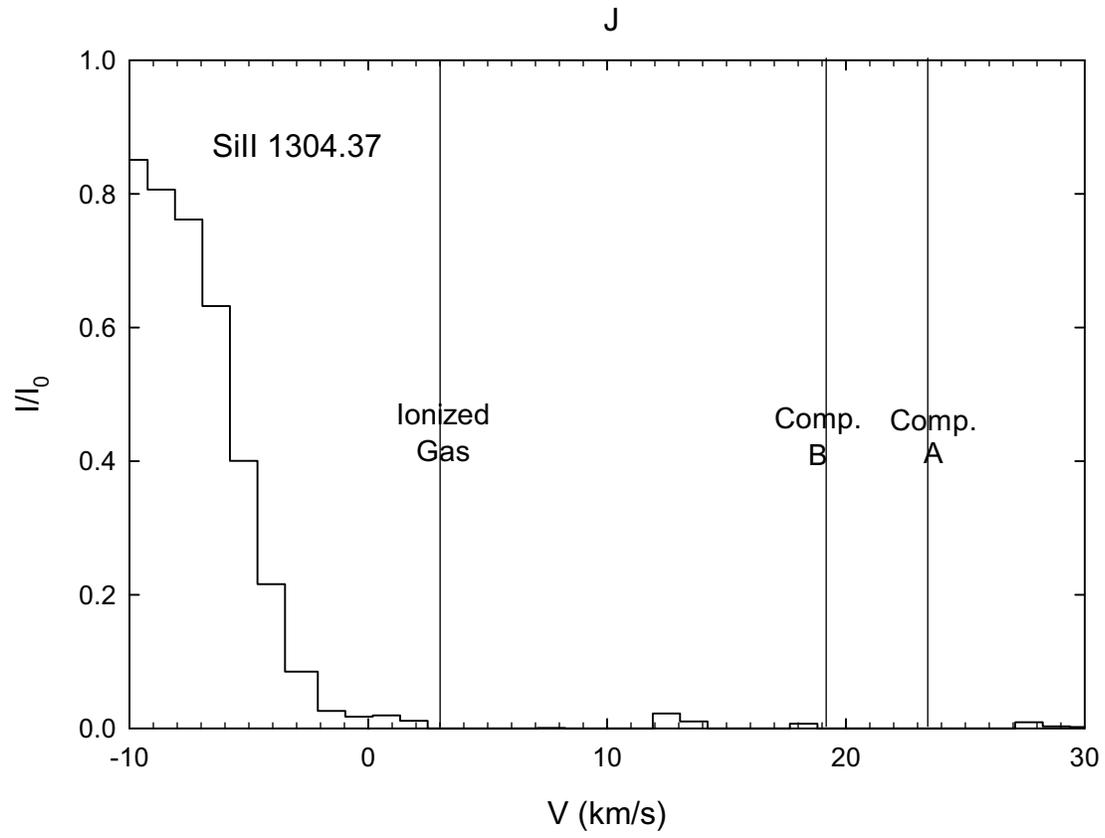

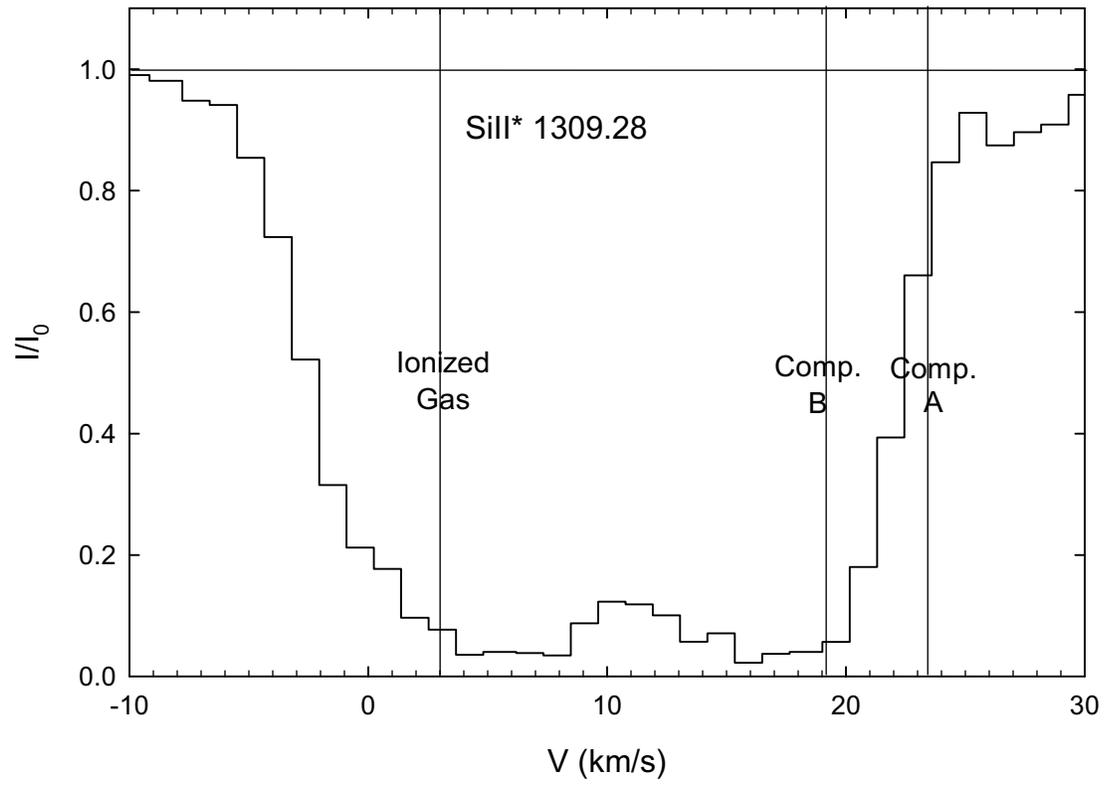

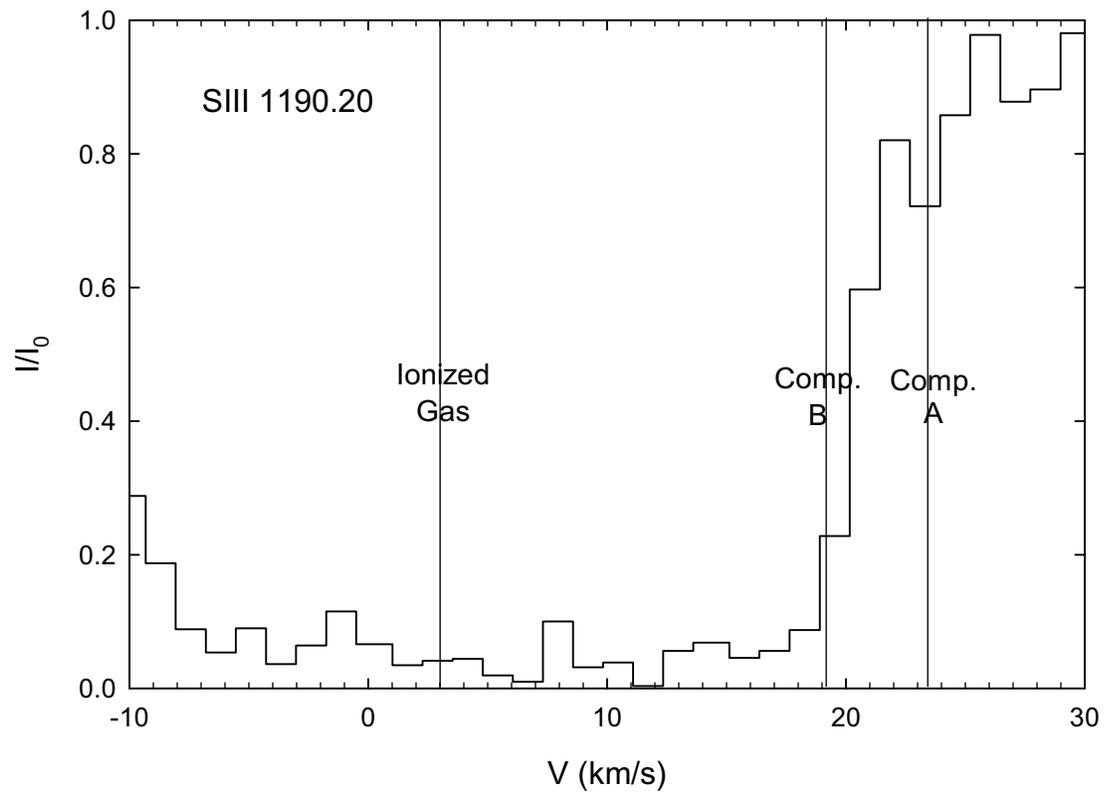

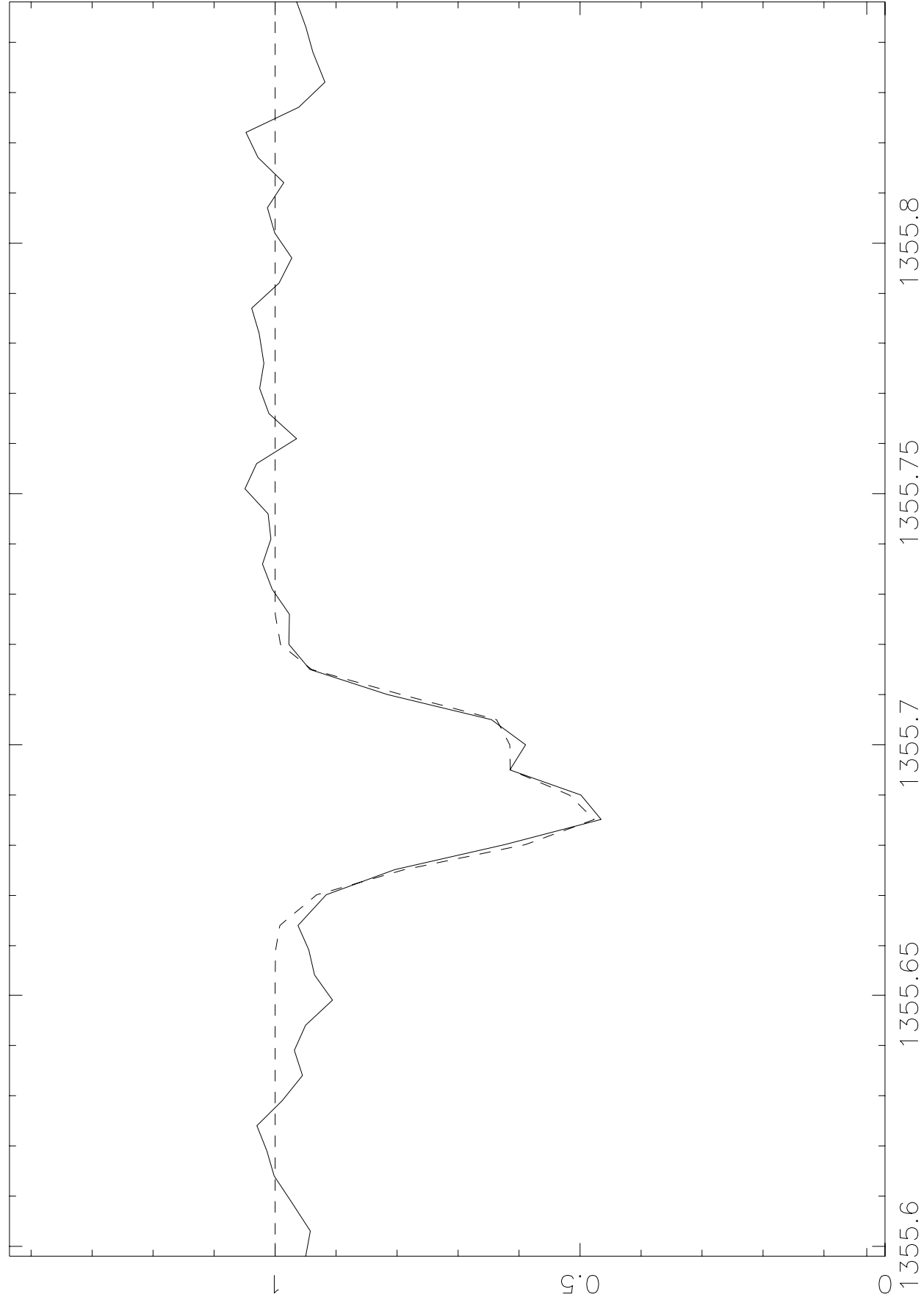

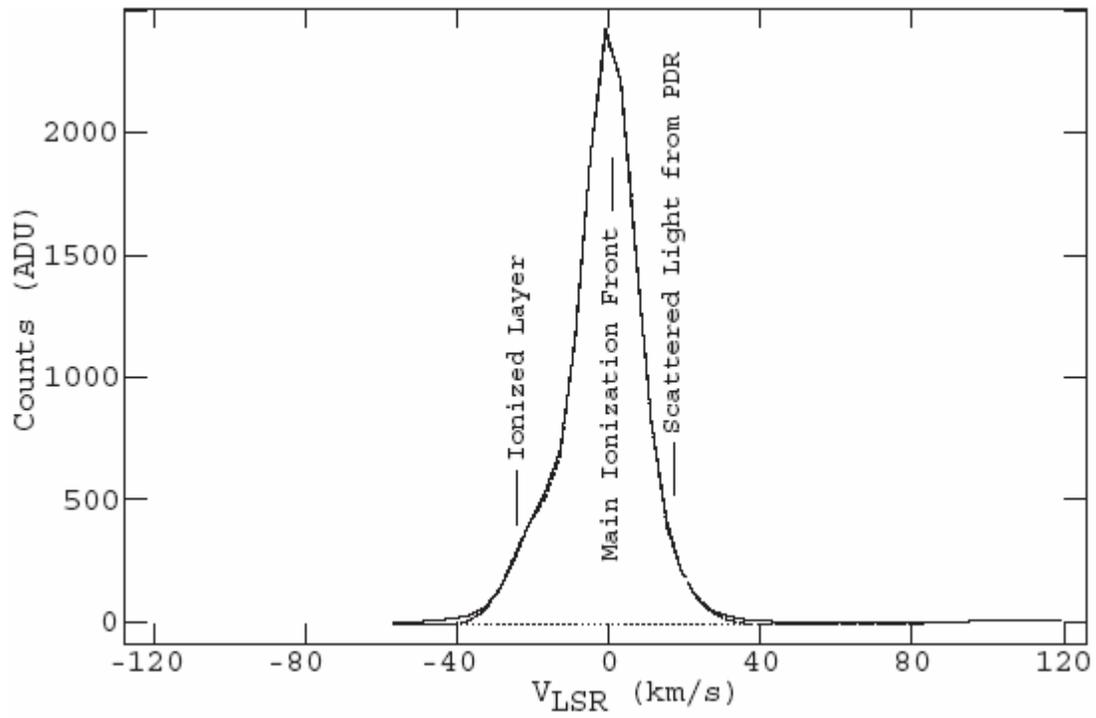

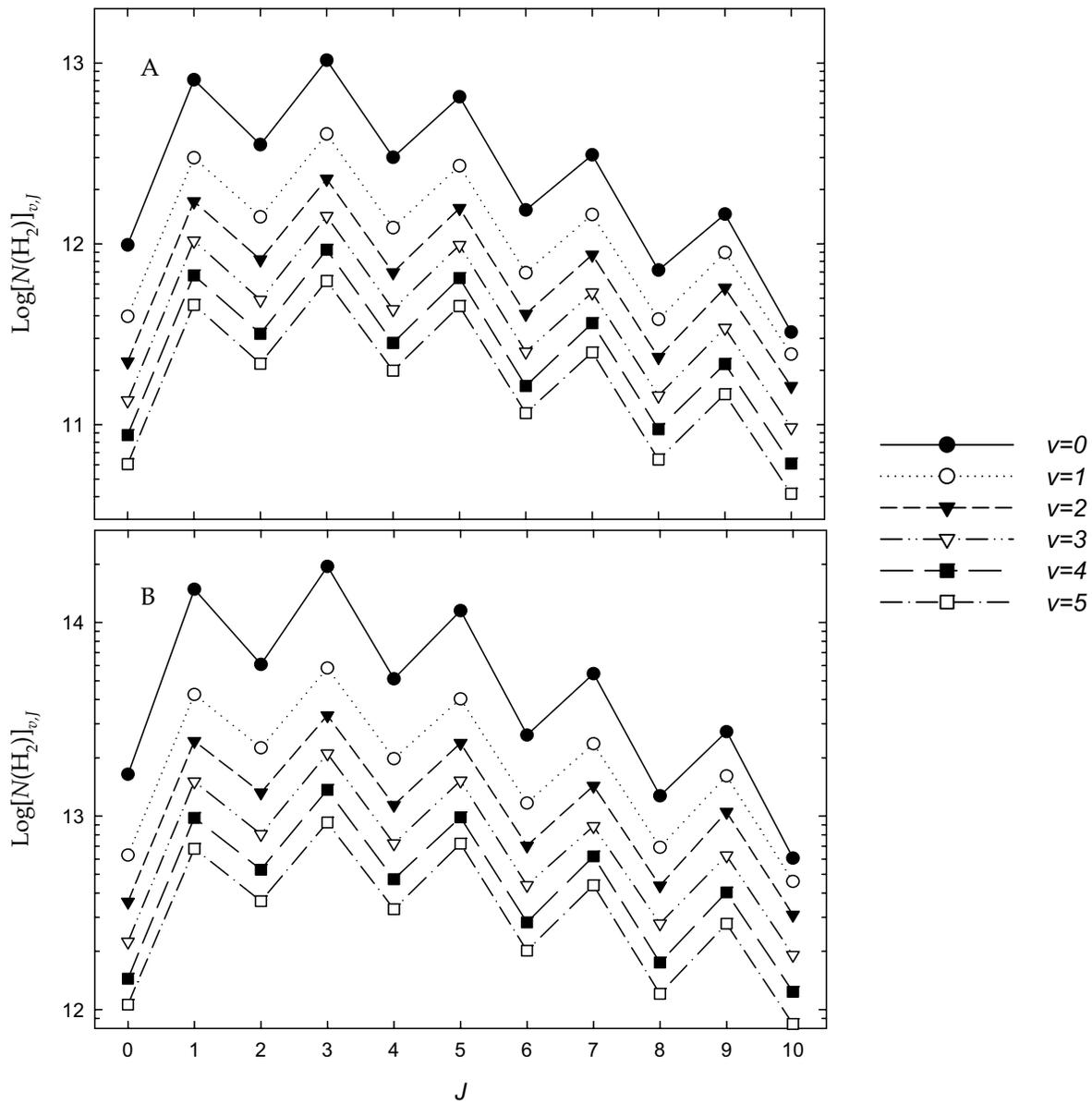

# 9 Figures

Figure 1  Line profiles toward θ$^1$ Ori B.  The x-axis is given in $V_{LSR}$, to allow for comparison between the H I (A) and UV (B-L) absorption lines.  The solid vertical lines represent the position of the 21 cm velocity components.  Most important to our analysis are the unsaturated absorption lines that correlate with the 21 cm absorption data.   Those absorption lines shown here are CI (B), CI* (C), CI**(D), OI(E), and two H$_2$ lines (H and I).  Additionally, we find S III absorption (L), which clearly must arise from an H$^+$ region.  There also appears to be some overlap in velocity space between S III and Si II (J), Si II*(K), which indicates at least a portion of this absorption is coming from the ionized gas.

Figure 2  Normalized STIS spectra of the O I] 1355.5977Å absorption line.  The S/N for this spectrum is 30-35.  Two components are clearly seen in the data.  The dotted line shows the fit to the data, using VPFIT.

Figure 3 The profile of the [N II] emission line of a region to the southeast of the Trapezium from the data described in Doi, O'Dell, & Hartigan (2004) is shown, together with its deconvolution into three major velocity components as described in the text.  The featured designated as the "Ionized Layer" corresponds to the newly identified H II layer that produces the newly discovered S III and P III UV absorption lines and the He I 3889Å line that has been observed for half a century but never explained.

Figure 4 Predicted H$_2$ column densities for each component from the A04 model and $n_H$ derived in this work.  Due to the proximity of the Veil to the Trapezium, UV pumping of H$_2$ forces the level populations out of thermodynamic equilibrium.  For a given $v$ the largest column densities are in the $J$=1 and 3 states, which agrees with observation.